\documentclass[twocolumn,showpacs,preprintnumbers,amsmath,amssymb,superscriptaddress]{revtex4}%
\usepackage{graphicx}
\usepackage{dcolumn}
\usepackage{bm}
\usepackage{amsmath}
\usepackage{amsfonts}
\usepackage{amssymb}%
\begin{document}
\title{Entanglement dynamics of a superconducting phase qubit coupled to a two-level system}
\author{Guozhu Sun}
\email{gzsun@nju.edu.cn}
\affiliation{Research Institute of Superconductor Electronics, School of Electronic Science
and Engineering, Nanjing University, Nanjing 210093, China}
\author{Zhongyuan Zhou}
\affiliation{Department of Chemistry, University of Kansas, Lawrence, KS 66045, USA}
\author{Bo Mao}
\affiliation{Department of Physics and Astronomy, University of Kansas, Lawrence, KS 66045, USA}
\author{Xueda Wen}
\affiliation{Department of Physics, University of Illinois at Urbana-Champaign, Urbana, IL
61801, USA}
\author{Peiheng Wu}
\affiliation{Research Institute of Superconductor Electronics, School of Electronic Science
and Engineering, Nanjing University, Nanjing 210093, China}
\author{Siyuan Han}
\email{han@ku.edu}
\affiliation{Department of Physics and Astronomy, University of Kansas, Lawrence, KS 66045, USA}
\date{\today}

\begin{abstract}
We report the observation and quantitative characterization of driven and
spontaneous oscillations of quantum entanglement, as measured by concurrence,
in a bipartite system consisting of a macroscopic Josephson phase qubit
coupled to a microscopic two-level system. The data clearly show the
behavior of entanglement dynamics such as sudden death and revival,
and the effect of decoherence and ac driving on entanglement.

\end{abstract}
\maketitle

\section{INTRODUCTION}
Entanglement is a unique property manifesting quantum correlation of
multiparticle quantum systems that has no classical counterpart. It has been
one of the most fascinating and nonintuitive concepts of quantum mechanics
and has stimulated extensive debate\cite{Schrodinger,PhysRev.47.777}. Recently, interest in entanglement has
intensified since it is considered as one of the key resources for quantum
information processing \cite{NatureCommunicationsGuo,PhysRevLett.106.257002} and as a consequence
a variety of properties of entanglement have been discovered
\cite{PhysicsReports.415.207,amico:517,horodecki:865}.
Nevertheless, many fundamental questions about entanglement remain open,
including the entanglement of autonomous open quantum systems, the effect of
external driving on entanglement, and the mechanism of damped entanglement oscillation (DEO), entanglement sudden death (ESD) and ESD revival (ESDR). Another important issue in the experimental study of
entanglement dynamics is to find simple methods to measure entanglement.

Entanglement can exist not only in microscopic but also in macroscopic systems
such as Josephson phase qubits (JPQs) \cite{Nature.467.570,Nature.467.574},
which are basically current or flux biased Josephson tunnel junctions having
Josephson coupling energy $E_{J}$ much greater than charging energy $E_{c}$.
JPQs are essentially manufacturable atoms
whose Hamiltonians can be custom designed and realized with integrated circuit
fabrication technology \cite{RevModPhys.73.357,PhysToday,Nature.453.1031}. This unique property makes the JPQ a good test-bed for studying
fundamental issues in quantum mechanics and a promising candidate for
implementing quantum information processing. In our experiment reported here a
flux biased JPQ, which is a radio frequency superconducting quantum
interference device consisting of a superconducting loop of inductance
$L\approx770$ pH interrupted by a $4.8$ $\mu$m$^{2}$ Josephson junction of
capacitance $C\approx240$ fF and critical current $I_{c}\approx1.4$ $\mu$A, is
used as shown in Fig. 1(a). The two lowest levels in the upper well of the
strongly tilted double well potential form the two computational basis states
$|0\rangle$ and $|1\rangle$. The energy level spacing, $\omega_{10}$ (for
convenience we set $\hbar\equiv h/2\pi=1$ where $h$ is Planck's constant),
between the two basis states can be continuously tuned by varying the amount
of magnetic flux inductively coupled to the superconducting loop.
Although atomic size defects in tunnel barrier
of the Josephson junction, which are essentially microscopic two-level systems
(TLSs), are considered as one of the major sources of energy relaxation and
decoherence in JQP \cite{PhysRevLett.93.077003,PhysRevLett.95.210503}, they
also have the potential to be utilized as a useful resource for quantum
information processing
\cite{PhysRevLett.97.077001,NaturePhysics.4.523,Nat.Commun., NJP2,PhysRevB.83.180507}. In this work, we
used a TLS coupled to a JPQ to investigate the dynamics of entanglement of the
coupled bipartite system. We show that for a bipartite system described by
Hamiltonian (\ref{Hamil}) the degree of entanglement, quantified by
\textquotedblleft concurrence\textquotedblright \ \cite{PhysRevLett.80.2245},
in both driven and free evolution states can be obtained by measuring the
state of JPQ alone, which is quite different from the conventional tomography
method in studying the entanglement. The results clearly show that resonant ac drive
and always-on qubit-TLS interaction together generates entanglement
oscillations and that in the subsequent free evolution the system may exhibit
DEO, ESD, and ESDR depending on the specific
bipartite states at the time of turning off the ac drive. A comparison between
the experimental results obtained with finite relaxation and decoherence and
the predictions based on analytical solution of the corresponding pure states
(i.e., relaxation and decoherence free) clearly shows that not only the
system-environment interaction but also the bipartite system's initial states
play important roles in determining the dynamic behavior of this type of open
quantum systems.

\section{EXPERIMENT}
\textrm{Fig. 1(b)} shows the schematics of the circuitry used. To distinguish
qubit states from those of TLS, we denote the ground and excited states of the
qubit (TLS) as $|0\rangle$ $(|g\rangle)$ and $|1\rangle$ $(|e\rangle)$,
respectively. The coupled qubit-TLS in a weak microwave field is described by
a four-level coupled bipartite system whose effective Hamiltonian, in the
basis of the four product states $\left\{  \left\vert 0g\right\rangle
,\left\vert 1g\right\rangle ,\left\vert 0e\right\rangle ,\left\vert
1e\right\rangle \right\}  $, is%
\begin{equation}
H=\left(
\begin{smallmatrix}
E_{0g} & \Omega_{\text{m}}\cos\omega t & 0 & 0\\
\Omega_{\text{m}}\cos\omega t & E_{1g} & g & 0\\
0 & g & E_{0e} & \Omega_{\text{m}}\cos\omega t\\
0 & 0 & \Omega_{\text{m}}\cos\omega t & E_{1e}%
\end{smallmatrix}
\right)  , \label{Hamil}%
\end{equation}
where the diagonal elements are the energies of corresponding product states,
$\Omega_{\text{m}}$ is the Rabi frequency of the JPQ, $g$ characterizes the
qubit-TLS coupling strength, and $\omega$ is the frequency of the microwave
field. Notice that at the degeneracy point one has $E_{1g}=E_{0e}$.The
system's parameters in Hamiltonian (\ref{Hamil}) were determined from
spectroscopy ($2g/2\pi=76.0$ $\pm1.2$ MHz), Rabi oscillation ($\Omega
_{\text{m}}/2\pi=63.0$ MHz), pump-probe ($T_{1}^{q}\approx61$ ns), and
pump-SWAP-delay-SWAP-probe ($T_{1}^{TLS}\approx146$ ns) experiments
\cite{Nat.Commun.} as shown in \textrm{Fig. 2}.

To measure dynamics of the coupled system, we first prepare the system in the
ground state $|0g\rangle$ at $t\leq0$, followed by applying a resonant
microwave pulse of width $t_{\text{mw}}$ at $t=0$ to coherently transfer the
system from $|0g\rangle$ to other states through qubit-microwave coupling
$\Omega_{\text{m}}$ and qubit-TLS coupling $g.$ After the pulse is terminated
at $t=t_{\text{mw}}$, the probability of finding the qubit in the state
$|1\rangle$, $P_{1}=P_{1g}+P_{1e},$ is measured after a time $t_{\text{free}}$
is elapsed from the end of the microwave pulse as shown in Fig. 1(c). The
procedure is repeated for different $t_{\text{mw}}$ and $t_{\text{free}}$ to
obtain $P_{1}$ as a function of $t_{\text{mw}}$ and free evolution time
$t_{\text{free}}$.

\section{RESULTS AND DISCUSSION}
To facilitate a direct comparison to analytical result in our experiment the
qubit-TLS was biased at the center of the anticrossing where $E_{1g}=E_{0e}$.
\textrm{Fig. 3(a)} shows the complete set of $P_{1}(t_{\text{mw}%
},t_{\text{free}})$ data measured. To clearly illustrate the effect of
$t_{\text{mw}}$ on vacuum Rabi oscillation of $P_{1}$ at $t\geq t_{\text{mw}%
},$ data taken with $t_{\text{mw}}=7.0,13.5,20.5$, and $27.0$ ns are plotted
as symbols in \textrm{Fig. 3(b)}. Notice that while the initial phase of
$P_{1}$ oscillation depends on $t_{\text{mw}}$ the frequency is independent of
$t_{\text{mw}}$ and its value, $76$ MHz, agrees very well with the size of the
splitting $2g$ obtained from the spectroscopic measurement. Furthermore, the data with
$t_{\text{mw}}=13.5$ ns and $t_{\text{mw}}=27.0$ ns show very little
oscillation while in a stark contrast the data with $t_{\text{mw}}=7.0$ ns and
$t_{\text{mw}}=20.5$ ns oscillate with much larger amplitudes. Their oscillations are
out of phase with each other indicating the importance of initial phase of vacuum Rabi oscillation in determining concurrence as discussed below.

To understand the dynamics of this coupled bipartite quantum system, we first
discuss the ideal situation of pure state evolution by solving the problem
analytically. In rotating frame Hamiltonian (1) is transformed to:
\begin{equation}
H_{r}=\left(
\begin{smallmatrix}
0 & \frac{\Omega_{\text{m}}}{2}\theta(t_{\text{mw}}-t) & 0 & 0\\
\frac{\Omega_{\text{m}}}{2}\theta(t_{\text{mw}}-t) & 0 & g & 0\\
0 & g & 0 & \frac{\Omega_{\text{m}}}{2}\theta(t_{\text{mw}}-t)\\
0 & 0 & \frac{\Omega_{\text{m}}}{2}\theta(t_{\text{mw}}-t) & 0
\end{smallmatrix}
\right)  , \label{simple}%
\end{equation}
where $\theta(t_{\text{mw}}-t)$ is the heaviside step function with
$\theta(t_{\text{mw}}-t)=1$ for $t<t_{\text{mw}}$ and $\theta(t_{\text{mw}%
}-t)=0$ for $t>t_{\text{mw}}$. It is obvious from Hamiltonian (\ref{simple})
that the dynamics occurs only in the subspace $\{|1g\rangle,|0e\rangle\}$ for
$t>t_{\text{mw}}$. Writing the wavefunction of the system in the form
$|\Psi(t)\rangle=c_{0g}(t)|0g\rangle+c_{1g}(t)|1g\rangle+c_{0e}(t)|0e\rangle
+c_{1e}(t)|1e\rangle$, we obtain probability amplitudes by solving the time
dependent Schr\"{o}dinger equation directly. For $t\leqslant t_{\text{mw}}$,
we have
\begin{align}
c_{0g}(t)  &  =\frac{1}{\Omega_{\text{m}}^{2}+\Omega_{+}^{2}}\left[
\Omega_{\text{m}}^{2}\cos(\Omega_{+}t/2)+\Omega_{+}^{2}\cos(\Omega
_{-}t/2)\right],\nonumber\\
c_{1g}(t)  &  =-i\frac{\Omega_{m}\Omega_{+}}{\Omega_{\text{m}}^{2}+\Omega
_{+}^{2}}[\sin(\Omega_{+}t/2)+\sin(\Omega_{-}t/2)]\nonumber\\
&  =-i\frac{2\Omega_{\text{m}}\Omega_{+}}{\Omega_{\text{m}}^{2}+\Omega_{+}%
^{2}}\sin\frac{\Omega_{\text{s}}t}{2}\cos\frac{gt}{2},\nonumber\\
c_{0e}(t)  &  =\frac{\Omega_{\text{m}}\Omega_{+}}{\Omega_{\text{m}}^{2}%
+\Omega_{+}^{2}}[\cos(\Omega_{+}t/2)-\cos(\Omega_{-}t/2)]\label{c_i}\\
&  =-\frac{2\Omega_{\text{m}}\Omega_{+}}{\Omega_{\text{m}}^{2}+\Omega_{+}^{2}%
}\sin\frac{\Omega_{\text{s}}t}{2}\sin\frac{gt}{2},\nonumber\\
c_{1e}(t)  &  =\frac{-i}{\Omega_{\text{m}}^{2}+\Omega_{+}^{2}}\left[
\Omega_{\text{m}}^{2}\sin(\Omega_{+}t/2)-\Omega_{+}^{2}\sin(\Omega
_{-}t/2)\right]  ,\nonumber
\end{align}
with $\Omega_{+}=\Omega_{\text{s}}+g$, $\Omega_{-}=\Omega_{\text{s}}-g$, and
$\Omega_{\text{s}}=\sqrt{\Omega_{\text{m}}^{2}+g^{2}}$. Notice that the
quantity measured directly in our experiment, $P_{1}(t)=|c_{1g}(t)|^{2}%
+|c_{1e}(t)|^{2},$ undergoes anomalous Rabi oscillation which in general
contains all three frequency components $\Omega_{+}$, $\Omega_{-}$, and
$\Omega_{\text{s}}$.

When the system is driven by a resonant microwave field $P_{2}(t<t_{\text{mw}%
})$, being the probability of finding the system in the
subspace spanned by $|1g\rangle$ and $|0e\rangle$, undergoes sinusoidal oscillation:
\begin{equation}
P_{2}(t\leq t_{\text{mw}})=|c_{1g}|^{2}+|c_{0e}|^{2}=\frac{2\Omega_{\text{m}%
}^{2}\Omega_{+}^{2}}{(\Omega_{\text{m}}^{2}+\Omega_{+}^{2})^{2}}(1-\cos
\Omega_{\text{s}}t), \label{P2}%
\end{equation}
which shows that amplitude of the oscillation depends only on the ratio $g/\Omega
_{\text{m}}$. In our experiment, $g/\Omega_{\text{m}}\simeq0.60$ yielding
$P_{2\max}\simeq0.74$ which agrees well with the experiment.

For $t>t_{\text{mw}}$, $t_{\text{free}}=t-t_{\text{mw}}$, it is
straightforward to show%
\begin{align*}
c_{0g}(t)  &  =c_{0g}(t_{\text{mw}})\\
c_{1g}(t)  &  =\cos gt_{\text{free}}\cdot c_{1g}(t_{\text{mw}})-i\sin
gt_{\text{free}}\cdot c_{0e}(t_{\text{mw}})\\
c_{0e}(t)  &  =\cos gt_{\text{free}}\cdot c_{0e}(t_{\text{mw}})-i\sin
gt_{\text{free}}\cdot c_{1g}(t_{\text{mw}})\\
c_{1e}(t)  &  =c_{1e}(t_{\text{mw}}).
\end{align*}

In this case, the probability of finding the qubit in sate $|1\rangle$ can be
expressed as
\begin{equation}%
\begin{split}
P_{1}(t  &  >t_{\text{mw}})=|c_{\text{1e}}|^{2}+|c_{\text{1g}}|^{2}\\
&  =P_{1e}(t_{\text{mw}})+\frac{1}{2}P_{2}(t_{\text{mw}})(1+\cos
(2gt_{\text{free}}+gt_{\text{mw}})),
\end{split}
\label{P1}%
\end{equation}
with $P_{1e}(t_{\text{mw}})=|c_{1e}(t_{\text{mw}})|^{2}$ being the population
of $|1e\rangle$ and $P_{2}(t_{\text{mw}})\equiv P_{1g}(t_{\text{mw}}%
)+P_{0e}(t_{\text{mw}})=\frac{4\Omega_{\text{m}}^{2}\Omega_{+}^{2}}%
{(\Omega_{\text{m}}^{2}+\Omega_{+}^{2})^{2}}\sin^{2}\frac{\Omega_{\text{s}%
}t_{\text{mw}}}{2}$ being the probability of finding the system in the
subspace spanned by $|1g\rangle$ and $|0e\rangle$ at $t=t_{\text{mw}}$ which
remains constant for $t>t_{\text{mw}}$. Eq. (\ref{P1}) shows that after
microwave is turned off the system undergoes vacuum Rabi oscillation caused by
the interaction between $|1g\rangle$ and $|0e\rangle.$ The angular frequency,
depth (peak-to-peak), initial phase, and bottom envelope of the oscillation
are $2g$, $P_{2}(t_{\text{mw}})$, $gt_{\text{mw}}$, and $P_{1e}(t_{\text{mw}%
})$, respectively. In addition, because $P_{0g}(t_{\text{mw}})+P_{2}%
(t_{\text{mw}})+P_{1e}(t_{\text{mw}})=1$ the difference between unity and the
top envelope of $P_{1}(t_{\text{mw}},t_{\text{free}})$ is just $P_{0g}%
(t_{\text{mw}})$.

Interestingly, when $\Omega_{\text{s}}t_{\text{mw}}=2n\pi$, $P_{2}%
(t_{\text{mw}})=0$, \textit{i.e.}, vacuum Rabi oscillation vanishes and one
has
\begin{equation}
P_{1}(t_{\text{mw}},t_{\text{free}})=\sin^{2}\frac{gt_{\text{mw}}}{2},
\end{equation}
which is independent of $t_{\text{free}}$. In addition, if $\Omega
_{+}t_{\text{mw}}\simeq(2k+1)\pi$, one has $P_{1}\simeq1$, which corresponds
to the population mainly occupying $|1e\rangle$; if $\Omega_{+}t_{\text{mw}%
}\simeq2k\pi$ , one has $P_{1}\simeq0$, which corresponds to the population
mainly occupying $|0g\rangle$. These two cases correspond to $t_{\text{mw}%
}=13.5$ ns and $27.0$ ns, respectively, as shown in \textrm{Fig. 3(b) }where
the exponential decay is due to energy relaxation.

Although the basic physics is the same, because of decoherence and relaxation
the experimental system investigated here cannot be represented by pure states
but mixed states described by the bipartite density operator $\mathbf{\rho
}\left(  t\right)  $. The diagonal matrix elements $\rho_{mm}$ and
off-diagonal matrix elements $\rho_{mn}$ $(m\neq n)$ represent the occupation
probability of the state $\left\vert m\right\rangle $ and coherence between
the states $\left\vert m\right\rangle $ and $\left\vert n\right\rangle $,
respectively. To simulate the system's dynamics, we solve the master equation
\cite{zhou1,Gardiner}%
\begin{equation}
\frac{d\rho_{mn}}{dt}=\sum_{m^{\prime}n^{\prime}}(-iL_{mn,m^{\prime}n^{\prime
}}+R_{mn,m^{\prime}n^{\prime}})\rho_{m^{\prime}n^{\prime}}, \label{rho}%
\end{equation}
where,\ $L_{mn,m^{\prime}n^{\prime}}=\left[  H_{mm^{\prime}}\delta_{n^{\prime
}n}-H_{n^{\prime}n}\delta_{mm^{\prime}}\right]  ,$ $H_{mm^{\prime}}$ is matrix
element of the system's Hamiltonian, and $R_{mn,m^{\prime}n^{\prime}}$ is the
damping rate matrix element whose value is proportional to the energy
relaxation rate \cite{zhou1}. Eq. (\ref{rho}) is numerically integrated to
obtain $\mathbf{\rho}\left(  t\right)  .$ The results is shown in \textrm{Fig.
3(c)} as well as the solid lines in \textrm{Fig. 2} (resonant ac drive)
and\textrm{ Fig. 3(b)} (free evolution). It can be seen clearly from
\textrm{Fig. 2} that unlike the normal sinusoidal Rabi oscillation observed in
the region of large qubit-TLS detuning, at the degeneracy point the
oscillation of $P_{1}(t\leq t_{\text{mw}})$ is clearly non-sinusoidal due to
more complicated dynamics of the driven four-level system
\cite{NJP_Ashhab,PhysRevB.80.172506,PhysRevB.81.100511,PhysRevB.82.132501}. In \textrm{Fig. 3(d)} we also
show the calculated $P_{1}$ based on the analytical solution Eq. (\ref{P1}) by
treating the effect of energy relaxation phenomenologically. Notice that
agreement between the experimental, numerical, and analytical results in the
entire range of driven and free evolution is very good confirming quantitative understanding of the system's dynamics.

To study the dynamics of qubit-TLS entanglement, we examine how concurrence,
denoted as $C_{\rho}$ \cite{PhysRevLett.80.2245} for mixed states, evolves with time. In \textrm{Fig.
4(a)} $C_{\rho}$ derived from the measured $P_{1}$ is shown. Of particular
interest is that $C_{\rho}$ is observed to undergo damped oscillation in both
of the driven and free run (\textit{i.e.}, autonomous) parts of the evolution.
\textrm{Fig. 4(c)} shows the Rabi-like oscillation when the system is driven
by the resonant microwave field. The corresponding $C_{\rho}$ (the dashed
line) oscillates and undergoes sudden death and revival repeatedly. The
extrema of $C_{\rho}$ are correlated strongly with the distinctive
\textquotedblleft shoulder\textquotedblright\ feature in $P_{1}.$ \textrm{Fig.
4(b)} shows $C_{\rho}$ in the free run part of the system's evolution with
$t_{\text{mw}}=7.0,13.5,20.5,$ and $27.0$ ns, respectively. When
$t_{\text{mw}}=7.0$ ns and $t_{\text{mw}}=20.5$ ns, the qubit and TLS are
mostly in $\{|1g\rangle$ and $|0e\rangle\}$. The coupling between these two
basis states via the $g(|1g\rangle\langle0e|+|0e\rangle\langle1g|)$ term of
the system Hamiltonian thus leads to time-dependent entanglement causing
$C_{\rho}$ to oscillate. In this case, the entanglement dynamics clearly
exhibits the phenomena of no ESD (NESD) and ESDR \cite{TingYu01302009}, respectively. In contrast, when $t_{\text{mw}}=27.0$ ns ($t_{\text{mw}%
}=13.5$ ns), the system is mostly in $|0g\rangle$ ($|1e\rangle$), which is
decoupled from all other three basis states after microwave was turned off
(neglecting energy relaxation). This results in a weak entanglement
characterized by a small overall value of concurrence $C_{\rho}$ at the start
of free evolution and, when decoherence is taken in account, the entanglement
sudden death.

Further insights into this type of system's entanglement dynamics and the
effects of environment can be obtained by examining time-dependent concurrence
of the corresponding pure state system and comparing it with the experimental
result. For the corresponding pure states, concurrence $\mathcal{C}%
=2|c_{0g}c_{1e}-c_{1g}c_{0e}|$ \cite{William2001}. From Eq. (\ref{c_i}), we obtain
\begin{equation}%
\begin{split}
\mathcal{C}  &  =|\frac{1}{(\Omega_{\text{m}}^{2}+\Omega_{+}^{2})^{2}}%
[\Omega_{\text{m}}^{4}\sin\Omega_{+}t_{\text{mw}}-\Omega_{+}^{4}\sin\Omega
_{-}t_{\text{mw}}\\
&  +2\Omega_{\text{m}}^{2}\Omega_{+}^{2}\sin gt_{\text{mw}}]+P_{2}%
(t_{\text{mw}})\sin(2gt_{\text{free}}+gt_{\text{mw}})|\\
&  \equiv|f(\Omega_{\text{m}},g,t_{\text{mw}})+P_{2}(t_{\text{mw}}%
)\sin(2gt_{\text{free}}+gt_{\text{mw}})|.
\end{split}
\label{concur}%
\end{equation}
Eq. (\ref{concur}) shows that when driven by a resonant microwave field
entanglement oscillation is rather complex which has three frequency
components $\Omega_{+},$ $\Omega_{-},$ and $g.$ In the free evolution stage
the so called preconcurrence \cite{PhysRevLett.80.2245}, which
is the quantity inside the absolute value sign of Eq. (\ref{concur}), undergoes sinusoidal oscillation with amplitude
$P_{2}(t_{\text{mw}}),$ which can be obtained directly from measured
$P_{1}(t_{\text{mw}},t_{\text{free}})$ according to Eq. (\ref{P1}), and a
vertical offset $|f(\Omega_{\text{m}},g,t_{\text{mw}})|$. Hence,
concurrence exhibits the "high-low" oscillation shown in Fig.4(b) unless offset of preconcurrence is zero. By varying
$t_{\text{mw}}$ and measuring the subsequent vacuum Rabi oscillation one can
trace time evolution of concurrence of the driven as well as the autonomous
stage of evolution. The result also shows that measuring the state of phase
qubit alone is sufficient to gain all information about entanglement
dynamics of this bipartite system.

In particular, when $\Omega_{\text{s}}t_{\text{mw}}=2n\pi$ one has
$P_{2}(t_{\text{mw}})=0$ as discussed above and thus only $|0g\rangle$ and $|1e\rangle$
contribute to concurrence:
\begin{equation}
\mathcal{C}=2|c_{0g}c_{1e}|=2\sqrt{P_{1}(1-P_{1})}=|\sin gt_{\text{mw}}|,
\end{equation}
which is independent of $t_{\text{free}}$. In addition,\textrm{ }when
$gt_{\text{mw}}\simeq(2k+1)\pi$ and $gt_{\text{mw}}\simeq2k\pi$, where $k$ is
an integer, we have $\mathcal{C}\ll1$ [notice the different vertical scales in
\textrm{Fig.4(b)}] because either $P_{1}\simeq1$ or $P_{1}\simeq0$.

According to Eq. (\ref{concur}), for pure states the amplitude of concurrence
oscillation remains constant for all $t$ and $\mathcal{C}$ reaches zero at a
discrete set of times only as shown in the insets of \textrm{Fig. 4(b)}. In
contrast, the experimental result displays a variety of interesting behaviors
including DEO, ESD, and ESDR illustrated in
\textrm{Fig. 4(b) }(from top to bottom). Because the main difference between
the measured qubit-TLS system and the corresponding hypothetical pure state
system is that the former is an open quantum system interacting with its
environment (e.g., energy relaxation and decoherence) while the latter is
isolated, our result shows that environment is the dominant mechanism of the
observed complex entanglement dynamics\cite{QuantumInfProcess.8.535}.

\section{CONCLUSION}
In summary, we have experimentally demonstrated that the coupled qubit-TLS
system is a test bed for quantitatively studying the dynamics of
bipartite entanglement. The measured time evolutions of this bipartite system,
either when driven by a resonant microwave field or in free evolution, agree
very well with the analytical and numerical solutions. Our results demonstrate
that in situations similar to those described here one not only can quantify
entanglement via concurrence by measuring the state of one constituent only,
but also be able to control the dynamics of the entanglement by adjusting the
interaction time between the qubit and the ac resonant driving field. A
comparison between the temporal evolutions of concurrence of the open and the
corresponding isolated systems indicates that for the bipartite system studied
here the entanglement oscillation and revival are originated from the qubit-TLS
coupling while the entanglement decay and sudden death are due to the coupling to the environment.

\begin{acknowledgments}
This work is partially supported by MOST (Grants No. 2011CB922104 and No. 2011CBA00200), NSFC (11074114,BK2010012), NSF Grant No. DMR-0325551. We acknowledge Northrop Grumman ES in Baltimore MD for technical and foundry support and thank R. Lewis, A. Pesetski, E. Folk, and J. Talvacchio for technical assistance.
\end{acknowledgments}

\newpage\begin{figure}[ptb]
\centering
\includegraphics[width=3.5in]{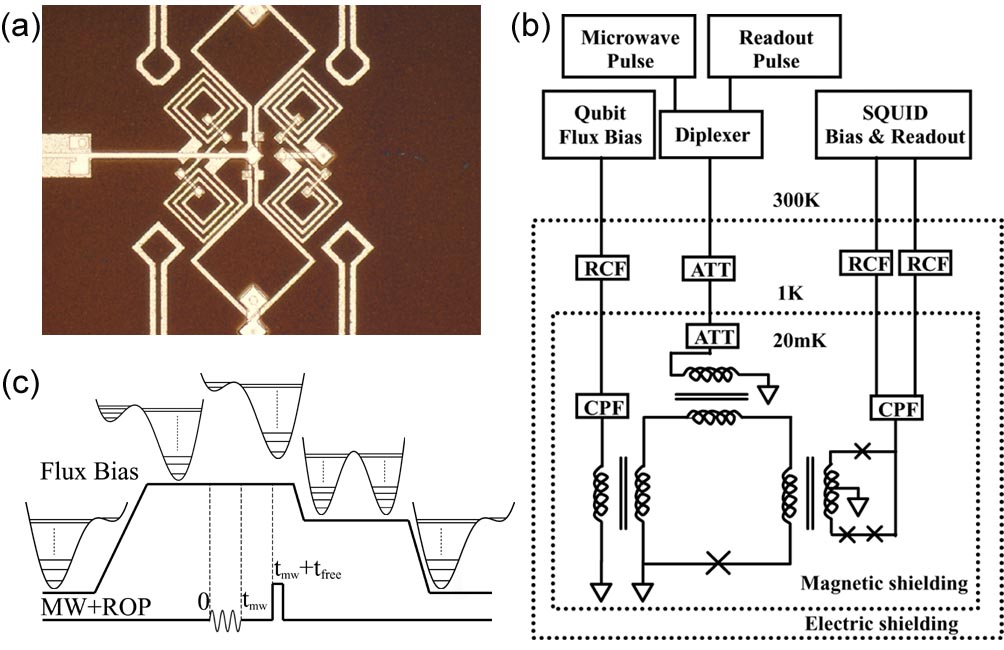}
\caption{(Color online)(a)
SEM of the sample which is fabricated by Al/AlOx/Al trilayers. (b) Schematic
of the qubit circuitry. Josephson junctions are denoted by the X symbols. (c)
A time profile of manipulation and measurement. The corresponding potential
energy landscape is also shown.}%
\label{fig:epsart}%
\end{figure}\begin{figure}[ptbptb]
\includegraphics[width=3.5in]{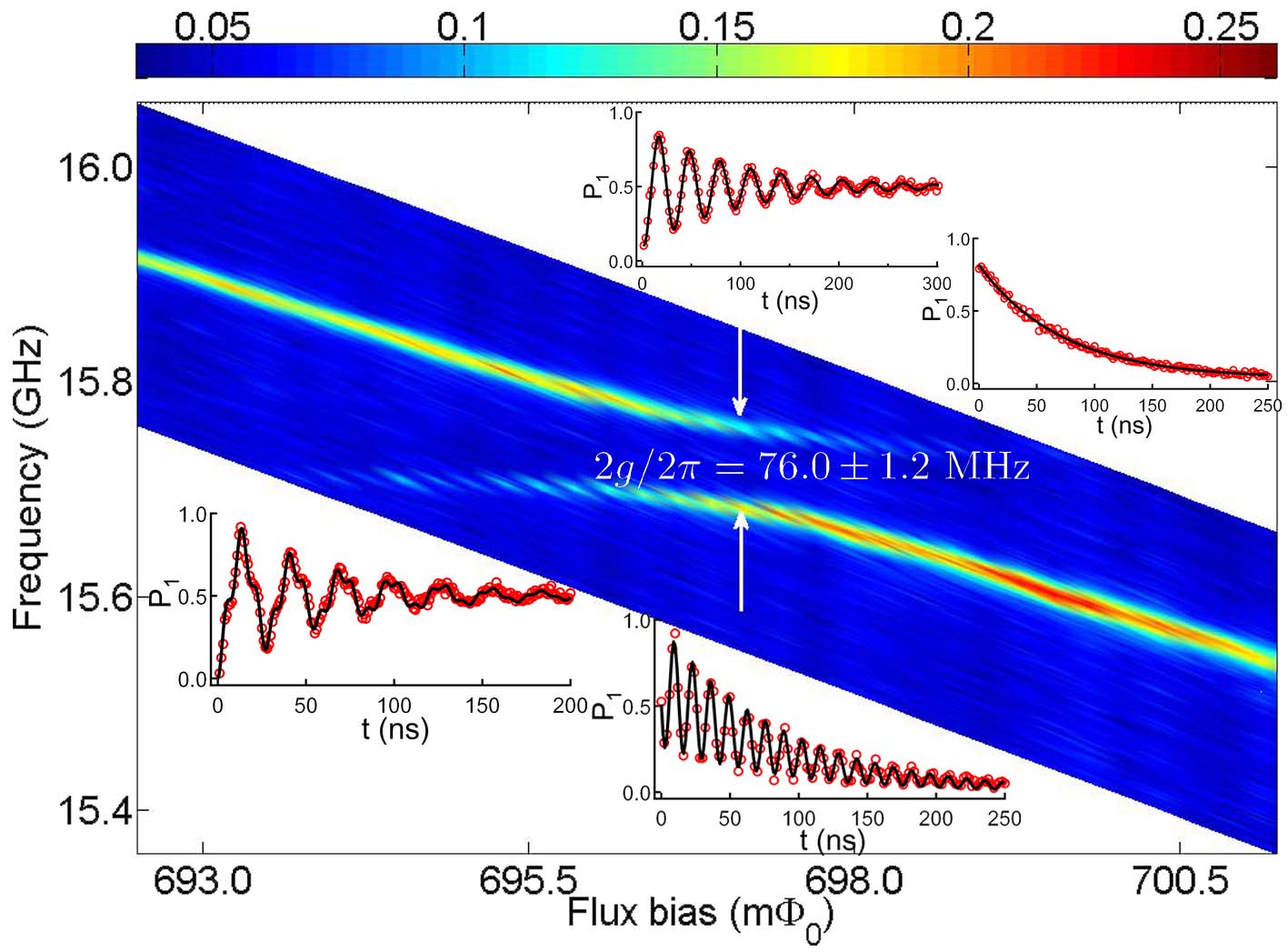}
\caption{(Color online)Spectroscopy
of the coupled qubit and TLS versus flux bias. The splitting due to the
coupling between the qubit and TLS is $2g/2\pi=76.0\pm1.2$ MHz at $f=15.722$ GHz.
The inserts are Rabi oscillation, $T_1$, Rabi beating and vacuum Rabi oscillation, separately (from top to bottom).}%
\label{fig:epsart}%
\end{figure}\begin{figure}[ptbptbptb]
\includegraphics[width=3.5in]{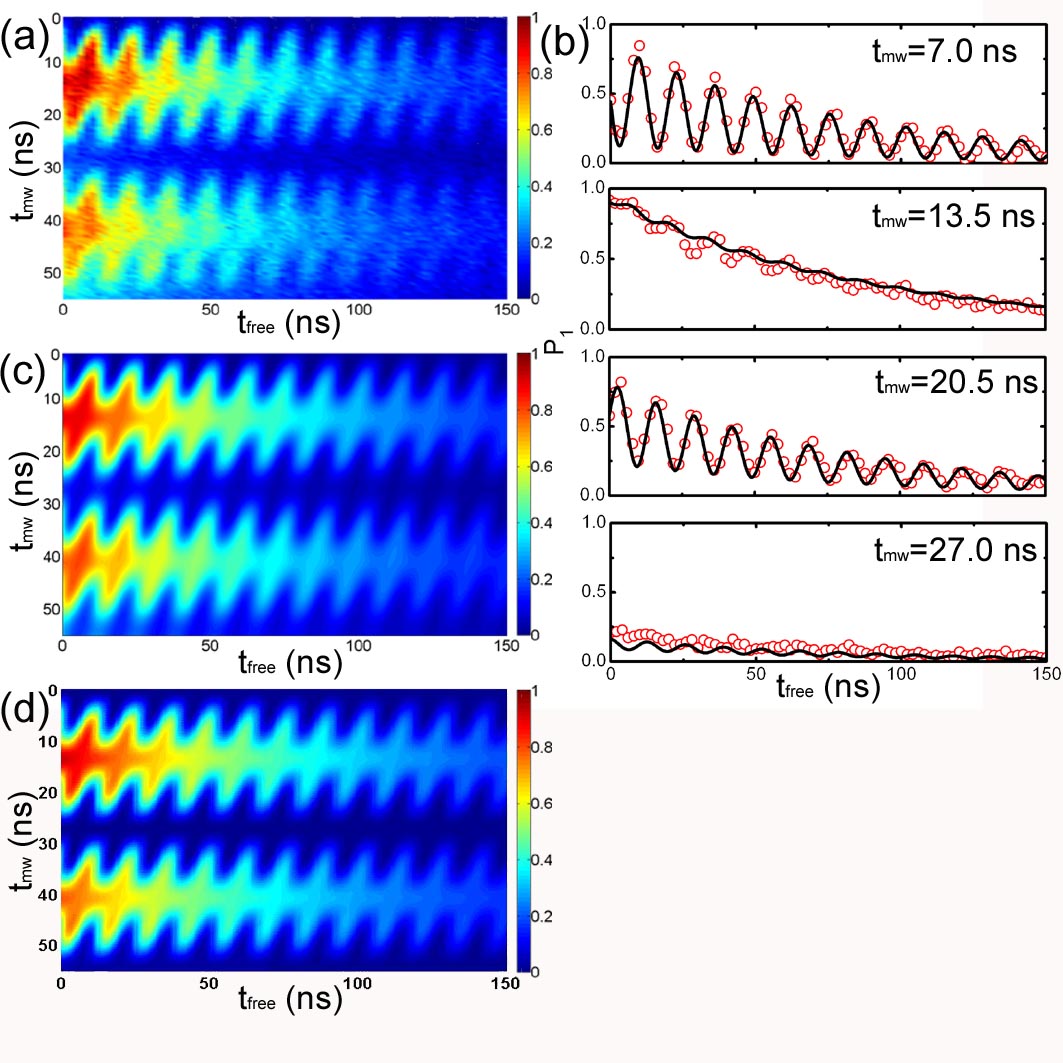}
\caption{(Color online)(a)
Experimentally measured $P_{1}$ versus $\mbox{t}_{\mathrm{mw}}$ and
$\mbox{t}_{\mathrm{free}}$. (b) $P_{1}$ oscillation with typical $\mbox{t}_{\mathrm{mw}}$. The symbols are the experimental measured $P_{1}$
and the lines are numerical result. (c) Numerical result of $P_{1}$ versus
$\mbox{t}_{\mathrm{mw}}$ and $\mbox{t}_{\mathrm{free}}$. (d) Analytical result
of $P_{1}$ versus $\mbox{t}_{\mathrm{mw}}$ and $\mbox{t}_{\mathrm{free}}$.}%
\label{fig:epsart}%
\end{figure}\begin{figure}[ptbptbptbptb]
\includegraphics[width=3.5in]{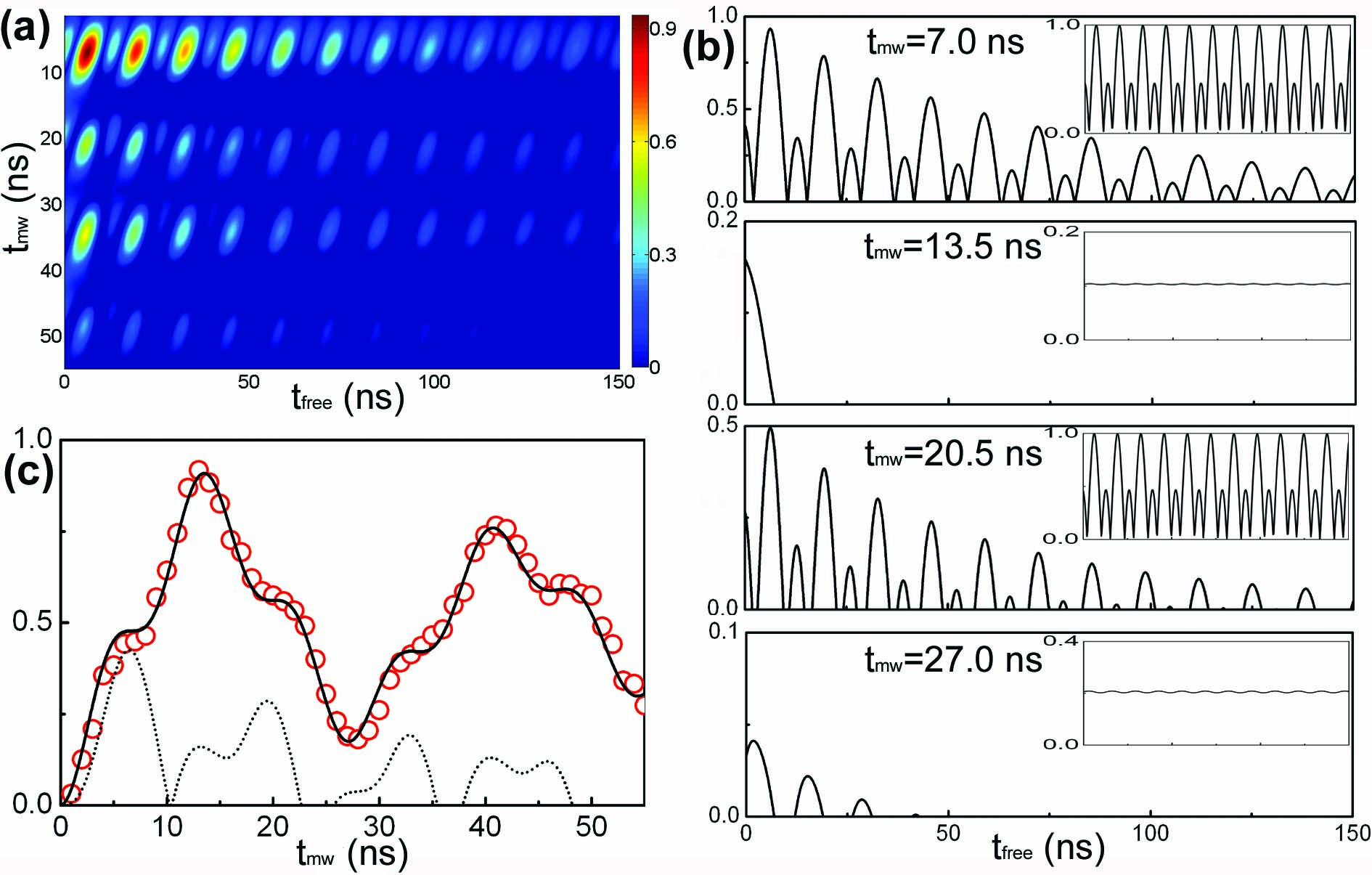}
\caption{(Color online)(a)
Concurrence as a function of $\mbox{t}_{\mathrm{mw}}$ and
$\mbox{t}_{\mathrm{free}}$ extracted from the experimental data. (b) Concurrence with
$\mbox{t}_{\mathrm{mw}}$=7.0, 13.5, 20.5, and 27.0 ns, respectively.
Analytical results of concurrence without decoherence are shown in the insets.
Notice that the different vertical scales are used. (c) $P_{1}$ and
concurrence as a function of $\mbox{t}_{\mathrm{mw}}$. Open circles are
experimental data and the solid line is the numerical result of $P_{1}$.
The dashed line is the corresponding concurrence.}%
\label{fig:epsart}%
\end{figure}


\begin{thebibliography}{1}

\bibitem{Schrodinger} E. Schr\"{o}dinger, Naturwissenschaften \textbf{23}, 807 (1935).
\bibitem{PhysRev.47.777} A. Einstein, B. Podolsky, B.  and N. Rosen, Phys. Rev. \textbf{47}, 777 (1935).
\bibitem{NatureCommunicationsGuo} J.-S. Xu, X.-Y. Xu, C.-F.Li, C.-J. Zhang, X.-B. Zou, and G.-C. Guo, Nature Communications \textbf{1}, 7 (2010).
\bibitem{PhysRevLett.106.257002} Y. Hu and L. Tian, Phys. Rev. Lett. \textbf{106}, 257002 (2011).
\bibitem{PhysicsReports.415.207} F. Minterta, A. R. R. Carvalhoa, M. Ku\'s and A. Buchleitner, Phys. Rep. \textbf{415}, 207 (2005).
\bibitem{amico:517} L. Amico, R. Fazio, A. Osterloh and V. Vedral, Rev. Mod. Phys. \textbf{80}, 517 (2008).
\bibitem{horodecki:865} R. Horodecki, P. Horodecki, M. Horodecki and K. Horodecki, Rev. Mod. Phys. \textbf{81}, 865 (2009).
\bibitem{Nature.467.570} M. Neeley, R. C. Bialczak, M. Lenander, E. Lucero, M. Mariantoni, A. D. O'Connell, D. Sank, H. Wang, M. Weides, J. Wenner, Y. Yin, T. Yamamoto, A. N. Cleland and J. M. Martinis, Nature \textbf{467}, 570 (2010).
\bibitem{Nature.467.574} L. DiCarlo, M. D. Reed, L. Sun, B. R. Johnson, J. M. Chow, J. M. Gambetta, L. Frunzio, S. M. Girvin, M. H. Devoret and R. J. Schoelkopf, Nature \textbf{467},574 (2010).
\bibitem{RevModPhys.73.357} Y. Makhlin, G. Sch\"on, A. Shnirman, Rev. Mod. Phys. \textbf{73}, 357 (2001).
\bibitem{PhysToday} J. You and F. Nori, Physics Today \textbf{58}(11), 42 (2005).
\bibitem{Nature.453.1031} J. Clarke and F. K. Wilhelm, Nature \textbf{453}, 1031 (2008).
\bibitem{PhysRevLett.93.077003} R. W. Simmonds, K. M. Lang, D. A. Hite, S. Nam, D. P. Pappas, and J. M. Martinis, Phys. Rev. Lett. \textbf{93}, 077003 (2004).
\bibitem{PhysRevLett.95.210503} J. M. Martinis, K. B. Cooper, R. McDermott, M. Steffen, M. Ansmann, K. D. Osborn,K. Cicak, S. Oh, D. P. Pappas, R. W. Simmonds, C. C. Yu, Phys. Rev. Lett. \textbf{95}, 210503 (2005).
\bibitem{PhysRevLett.97.077001} A. M. Zagoskin, S. Ashhab, J. R. Johansson and F. Nori \emph{et al}., Phys. Rev. Lett. \textbf{97}, 077001 (2006).
\bibitem{NaturePhysics.4.523} M. Neeley, M. Ansmann, R. C. Bialczak, M. Hofheinz, E. L. N. Katz, A. O'Connell, H. Wang, A. N. Cleland and J. M. Martinis, Nature Physics \textbf{4}, 523 (2008).
\bibitem{Nat.Commun.} G. Sun, X. Wen, B. Mao, J. Chen, Y. Yu, P. Wu and S. Han, Nature Communications \textbf{1}, 51 (2010).
\bibitem{NJP2} G. J. Grabovskij, P. Bushev, J. H. Cole, C. M\"uller, J. Lisenfeld, A. Lukashenko and A. V. Ustinov, New J. Phys. \textbf{13}, 063015 (2011).
\bibitem{PhysRevB.83.180507} G. Sun, X. Wen, B. Mao, Y. Yu, J. Chen, W. Xu, L. Kang, P. Wu and S. Han, Phys. Rev. B \textbf{83}, 180507(R) (2011).
\bibitem{PhysRevLett.80.2245} W. K. Wootters, Phys. Rev. Lett. \textbf{80}, 2245 (1998).
\bibitem{zhou1} Z. Zhou, S.-I. Chu and S. Han, J. Phys. B: At. Mol. Opt. Phys. \textbf{41}, 045506 (2008).
\bibitem{Gardiner} G. W. Gardiner and P. Zoller, Quantum Noise (Springer Verlag, Berlin, 2004), 3rd ed.
\bibitem{NJP_Ashhab} S. Ashhab, J. R. Johansson and F. Nori, New J. Phys. \textbf{8}, 103 (2006).
\bibitem{PhysRevB.80.172506} A. Lupa\ifmmode \mbox{\c{s}}\else \c{s}\fi{}cu, P. Bertet, E. F. C. Driessen, C. J. P. M. Harmans and J. E. Mooij, Phys. Rev. B \textbf{80}, 172506 (2009).
\bibitem{PhysRevB.81.100511} J. Lisenfeld, C. M\"uller, J. H. Cole, P. Bushev, A. Lukashenko, A. Shnirman, and A. V. Ustinov, Phys. Rev. B \textbf{81}, 100511(R) (2010).
\bibitem{PhysRevB.82.132501} G. Sun, X. Wen, B. Mao, Z. Zhou, Y. Yu, P. Wu, S. Han, Phys. Rev. B \textbf{82}, 132501 (2010).
\bibitem{TingYu01302009} T. Yu and J. H. Eberly, Science \textbf{323}, 598 (2009).
\bibitem{William2001} W. K. Wootters, Quantum Inf. and Compt. \textbf{1}, 27 (2001).
\bibitem{QuantumInfProcess.8.535} J. P. Paz and A. J. Roncaglia, Quantum Inf. Process. \textbf{8}, 535 (2009).
\end{thebibliography}
\end{document}